\documentclass[superscriptaddress,prc,twocolumn,showpacs,preprintnumbers,amsmath,amssymb,bibnotes,altaffilletter]{revtex4-1}
\usepackage{orcidlink}
\usepackage{centernot}
\usepackage{amsmath} 
\usepackage{hyperref}% add hypertext capabilities
\usepackage{graphicx}% Include figure files
\usepackage{lineno}
\def\DEM{{\sc Demonstrator}} %Demonstrator in small caps
\def\MJD{{\sc Majorana Demonstrator}}
\def\76Ge{${}^{76}$Ge}
\def\0vbb{$0\nu\beta\beta$}
\begin{document}

\title{Rare multi-nucleon decays with the full data sets of the \MJD}

\newcommand{\ITEP}{National Research Center ``Kurchatov Institute'', Kurchatov Complex of Theoretical and Experimental Physics, Moscow, 117218 Russia}
\newcommand{\JINR}{Joint Institute for Nuclear Research, Dubna, 141980 Russia} 
\newcommand{\lbnl}{Nuclear Science Division, Lawrence Berkeley National Laboratory, Berkeley, CA 94720, USA}
\newcommand{\lanl}{Los Alamos National Laboratory, Los Alamos, NM 87545, USA}
\newcommand{\uw}{Center for Experimental Nuclear Physics and Astrophysics, and Department of Physics, University of Washington, Seattle, WA 98195, USA}
\newcommand{\unc}{Department of Physics and Astronomy, University of North Carolina, Chapel Hill, NC 27514, USA}
\newcommand{\duke}{Department of Physics, Duke University, Durham, NC 27708, USA}
\newcommand{\ncsu}{Department of Physics, North Carolina State University, Raleigh, NC 27695, USA}	
\newcommand{\ornl}{Oak Ridge National Laboratory, Oak Ridge, TN 37830, USA}
\newcommand{\ou}{Research Center for Nuclear Physics, Osaka University, Ibaraki, Osaka 567-0047, Japan}
\newcommand{\pnnl}{Pacific Northwest National Laboratory, Richland, WA 99354, USA}
\newcommand{\sdsmt}{South Dakota Mines, Rapid City, SD 57701, USA}
\newcommand{\usc}{Department of Physics and Astronomy, University of South Carolina, Columbia, SC 29208, USA}
\newcommand{\usd}{Department of Physics, University of South Dakota, Vermillion, SD 57069, USA}  
\newcommand{\ut}{Department of Physics and Astronomy, University of Tennessee, Knoxville, TN 37916, USA}
\newcommand{\tunl}{Triangle Universities Nuclear Laboratory, Durham, NC 27708, USA}
\newcommand{\williams}{Physics Department, Williams College, Williamstown, MA 01267, USA}
\newcommand{\ciemat}{Centro de Investigaciones Energ\'{e}ticas, Medioambientales y Tecnol\'{o}gicas, CIEMAT 28040, Madrid, Spain}
\newcommand{\iu}{Center for Exploration of Energy and Matter, and Department of Physics, Indiana University, Bloomington, IN 47405, USA}
\newcommand{\ucsd}{Hal\i c{\i}o\u{g}lu Data Science Institute, Department of Physics, University of California San Diego, CA 92093, USA}

\author{I.J.~Arnquist}\affiliation{\pnnl} 
\author{F.T.~Avignone~III}\affiliation{\usc}\affiliation{\ornl}
\author{A.S.~Barabash\,\orcidlink{0000-0002-5130-0922}}\affiliation{\ITEP}
\author{E.~Blalock\,\orcidlink{0000-0001-5311-371X}}\affiliation{\ncsu}\affiliation{\tunl}
\author{B.~Bos\,\orcidlink{0009-0008-5828-1745}}\affiliation{\unc}\affiliation{\tunl}
\author{M.~Busch}\affiliation{\duke}\affiliation{\tunl}	
\author{Y.-D.~Chan}\affiliation{\lbnl}
\author{J.R.~Chapman\,\orcidlink{0009-0004-9815-2981}}\affiliation{\unc}\affiliation{\tunl} 
\author{C.D.~Christofferson\,\orcidlink{0009-0005-1842-9352}}\affiliation{\sdsmt} 
\author{P.-H.~Chu\,\orcidlink{0000-0003-1372-2910}}\email{pchu@lanl.gov} \affiliation{\lanl} 
\author{C.~Cuesta\,\orcidlink{0000-0003-1190-7233}}\affiliation{\ciemat}	
\author{J.A.~Detwiler\,\orcidlink{0000-0002-9050-4610}}\affiliation{\uw}	
\author{Yu.~Efremenko}\affiliation{\ut}\affiliation{\ornl}
\author{H.~Ejiri}\affiliation{\ou}
\author{S.R.~Elliott\,\orcidlink{0000-0001-9361-9870}}\affiliation{\lanl}
\author{N.~Fuad\,\orcidlink{0000-0002-5445-2534}}\affiliation{\iu} 
\author{G.K.~Giovanetti\,\orcidlink{0000-0002-3125-0550}}\affiliation{\williams}
\author{M.P.~Green\,\orcidlink{0000-0002-1958-8030}}\affiliation{\ncsu}\affiliation{\tunl}\affiliation{\ornl}   
\author{J.~Gruszko\,\orcidlink{0000-0002-3777-2237}}\affiliation{\unc}\affiliation{\tunl} 
\author{I.S.~Guinn\,\orcidlink{0000-0002-2424-3272}}\affiliation{\ornl} 
\author{V.E.~Guiseppe\,\orcidlink{0000-0002-0078-7101}}\affiliation{\ornl}	
\author{R.~Henning\,\orcidlink{0000-0001-8651-2960}}\affiliation{\unc}\affiliation{\tunl}
\author{E.W.~Hoppe\,\orcidlink{0000-0002-8171-7323}}\affiliation{\pnnl}
\author{R.T.~Kouzes\,\orcidlink{0000-0002-6639-4140}}\affiliation{\pnnl}
\author{A.~Li\,\orcidlink{0000-0002-4844-9339}}\affiliation{\ucsd} 
\author{R.~Massarczyk\,\orcidlink{0000-0001-8001-9235}}\affiliation{\lanl}		
\author{S.J.~Meijer\,\orcidlink{0000-0002-1366-0361}}\affiliation{\lanl}	
\author{L.S.~Paudel\,\orcidlink{0000-0003-3100-4074}}\affiliation{\usd} 
\author{W.~Pettus\,\orcidlink{0000-0003-4947-7400}}\affiliation{\iu}	
\author{A.W.P.~Poon\,\orcidlink{0000-0003-2684-6402}}\affiliation{\lbnl}
\author{D.C.~Radford}\affiliation{\ornl}
\author{A.L.~Reine\,\orcidlink{0000-0002-5900-8299}}\affiliation{\iu}	
\author{K.~Rielage\,\orcidlink{0000-0002-7392-7152}}\affiliation{\lanl}
\author{D.C.~Schaper\,\orcidlink{0000-0002-6219-650X}}\altaffiliation{Present address: Indiana University, Bloomington, IN 47405, USA}\affiliation{\lanl} 
\author{S.J.~Schleich\,\orcidlink{0000-0003-1878-9102}}\affiliation{\iu} 
\author{D.~Tedeschi\,\orcidlink{0000-0002-2999-5676}}\affiliation{\usc}
\author{R.L.~Varner\,\orcidlink{0000-0002-0477-7488}}\affiliation{\ornl}  
\author{S.~Vasilyev}\affiliation{\JINR}	
\author{S.L.~Watkins\,\orcidlink{0000-0003-0649-1923}}\altaffiliation{Present address: Pacific Northwest National Laboratory, Richland, WA 99354, USA}\affiliation{\lanl} 
\author{J.F.~Wilkerson\,\orcidlink{0000-0002-0342-0217}}\affiliation{\unc}\affiliation{\tunl}\affiliation{\ornl}    
\author{C.~Wiseman\,\orcidlink{0000-0002-4232-1326}}\affiliation{\uw}		
\author{C.-H.~Yu\,\orcidlink{0000-0002-9849-842X}}\affiliation{\ornl}

\collaboration{{\sc{Majorana}} Collaboration}
%\noaffiliation

\begin{abstract}
The \MJD~was an ultra-low-background experiment designed for neutrinoless double-beta decay (\0vbb) investigation in \76Ge. Located at the Sanford Underground Research Facility in Lead, South Dakota, the \DEM~utilized modular high-purity Ge detector arrays within shielded vacuum cryostats, operating deep underground. The arrays, with a capacity of up to 40.4 kg (27.2 kg enriched to $\sim 88\%$ in \76Ge), have accumulated the full data set, totaling 64.5 kg yr of enriched active exposure and 27.4 kg yr of exposure for natural detectors. Our updated search improves previously explored three-nucleon decay modes in Ge isotopes, setting new partial lifetime limits of $1.83\times10^{26}$ years  (90\% confidence level) for \76Ge($ppp$) $\rightarrow$ $^{73}$Cu e$^+\pi^+\pi^+$ and \76Ge($ppn$) $\rightarrow$ $^{73}$Zn e$^+\pi^+$. The partial lifetime limit for the fully inclusive tri-proton decay mode of \76Ge is found to be $2.1\times10^{25}$ yr. Furthermore, we have updated limits for corresponding multi-nucleon decays.
\end{abstract}


\maketitle

%\section{Introduction}
Baryons include stable particles like protons, as well as neutrons, which are stable within nuclei but decay outside them. In addition, there are other unstable particles with different quark combinations. While unstable baryons may decay into stable baryons, the conservation of the total baryon number is usually maintained during the process. Baryon number violation ($\centernot{B}$) is one of the three Sakharov conditions~\cite{Sakharov:1967dj} necessary to explain the observed matter-antimatter asymmetry in the universe. These topics are comprehensively studied in the literature~\cite{Babu:2013jba,Dev:2022jbf,FileviezPerez:2022ypk}. For example, certain theories beyond the Standard Model, such as Grand Unified Theories (GUTs), predict processes where baryon number conservation is violated, suggesting that protons are not stable and have a finite partial lifetime. The current constraint on the proton's partial lifetime is $1.6\times10^{34}$ yr~\cite{Super-Kamiokande:2016exg}. When considering $\Delta B = 2$ decays, indicating changes in baryon number by two units, the tightest constraints on its partial lifetime now approach approximately $10^{32}$ yr~\cite{Super-Kamiokande:2015pys,Super-Kamiokande:2015jbb}. 

Here we investigate the decay processes of three stable baryons, such as protons or neutrons, known as tri-nucleon decays. Tri-nucleon decays~\cite{Babu:2003qh} are studied within the framework of the Standard Model with small neutrino masses where the anomaly-free $Z_6$ symmetry allows $\Delta B = 3$ transitions, primarily due to a dimension 15 operator, while simultaneously prohibiting lower-order transitions such as $\Delta B = 2$ or $\Delta B = 1$. $Z_6$ is a discrete symmetry group of order 6, describing additional symmetries and charge assignments beyond the Standard Model~\cite{Bakker:2004}. In a $\Delta B = 3$ tri-nucleon decay, three baryons vanish from the nucleus, often leading to the formation of an unstable isotope. Previous investigations into $\Delta B = 3$ processes in Xe isotopes~\cite{Bernabei:2006tw,EXO-200:2017hwz} and $^{127}$I~\cite{Hazama:1994zz} primarily focused on uncovering invisible decay channels~\cite{Heeck:2019kgr}, assuming the initial tri-nucleon decay or its disappearance went unobserved. These studies concentrated on detecting evidence of the process solely through the decay of the unstable product. Recently, the \MJD~reported their initial findings on tri-nucleon decays involving germanium isotopes~\cite{Majorana:2018pdo}. The GERDA collaboration published a study on inclusive tri-nucleon decays in \76Ge~\cite{GERDA:2023uuw}, utilizing the isotope ${}^{73}$Ga decay to ${}^{73m}$Ge as a tracer. A process with a similar signature in the detectors has been observed in the \DEM~due to cosmogenic muon backgrounds~\cite{Majorana:2021lgr} that limit sensitivity to that tri-nucleon decay mode. 

Babu \textit{et al.}~\cite{Babu:2003qh} calculated potential decay modes for tri-nucleon decays with $\Delta B=3$, involving combinations of proton and neutron decays. These modes include $ppp\rightarrow e^+ \pi^+ \pi^+$, $ppn\rightarrow e^+ \pi^+$, $pnn\rightarrow e^+ \pi^0$, and $nnn\rightarrow \bar{\nu} \pi^0$. The presence of $e^+$ or $\bar{\nu}$ in the output provides a test for lepton number violation with $\Delta L=1$, whereas neutrinoless double beta decay corresponds to $\Delta L=2$.  We have also considered the decays with $\Delta B = 2$~\cite{Heeck:2019kgr,Proceedings:2020nzz}, such as $pp\rightarrow \pi^+ \pi^+$, $pn\rightarrow \pi^0 \pi^+$, and $nn\rightarrow \pi^+ \pi^-$, as well as $\pi^0\pi^0$. In this study, we present updated results for the corresponding $\Delta B = 2$ and 3 decays using the complete data set from the \DEM.

%\section{\MJD}
%Describe MJD and data set
The \MJD~\cite{Majorana:2013cem,Abgrall:2025tsj} was an experiment at the Sanford Underground Research Facility~\cite{Heise:2022iaf} probing neutrinoless double-beta decay (\0vbb). Housed underground in Lead, South Dakota, the \DEM~employed modular high-purity germanium detectors~\cite{Barbeau:2007qi} in shielded vacuum cryostats and a muon veto system~\cite{Majorana:2016ifg} to ensure an ultra-low-background environment~\cite{Hoppe:2014nva,Majorana:2016lsk}. The electronics system~\cite{Majorana:2021mtz} of the \DEM~enabled energy spectra measurement (1 keV to 10 MeV) with pulse-shape analysis~\cite{Majorana:2019ftu,Majorana:2020xvk} for signal-background discrimination. Advanced correction methods, including nonlinearity and charge trapping~\cite{Majorana:2020llj,Majorana:2022vai}, provided exceptional energy resolution ($2.52\pm 0.08$ keV at 2039 keV). The calibration system~\cite{Majorana:2017vdg} used $^{228}$Th sources around detectors for reliable response~\cite{Majorana:2023kdx}.

Throughout the data sets, the \DEM~utilized three distinct types of germanium detectors, each characterized by varying pulse-shape analysis efficiency. These detector types included P-type point-contact (PPC) detectors~\cite{Luke:1989,Barbeau:2007qi}, enriched to $87.4\pm 0.5\%$ in \76Ge (and $12.6\%$ $^{74}$Ge), with masses ranging from 0.6 to 1.2 kg. Additionally, the inverted coaxial point contact (ICPC) detectors~\cite{COOPER201125} enriched to $88\pm 1\%$ in \76Ge (and $12\%$ $^{74}$Ge) were also employed, with masses ranging from 1.3 to 2.1 kg. Finally, the \DEM~incorporated Broad Energy Germanium (BEGe) detectors~\cite{Canberra} made from natural germanium (20.5$\%$ $^{70}$Ge, 27.4$\%$ $^{72}$Ge, 7.8$\%$ $^{73}$Ge, 36.5$\%$ $^{74}$Ge, and 7.8$\%$ $^{76}$Ge), each weighing approximately 0.6 kg with the total exposure 27.4 kg yr.  With a combined exposure of 64.5 kg yr of enriched \76Ge  (61.64 kg yr for PPC and 2.82 kg yr for ICPC), the \DEM~established a half-life ($T_{1/2}$) limit for \0vbb~in \76Ge, with $T_{1/2}>8.3\times 10^{25}$ yr ($90\%$ confidence level)~\cite{Majorana:2022udl}. 
\begin{table*}
    \centering
    \begin{tabular}{c|c|c|c|c|c|c|c|c|c|c}
	\hline 
	\hline
 Decay Mode & $\epsilon_0$ &
 $\epsilon_{\tau_1}$ &
 $\epsilon_{E_1}$ &
 $\epsilon_{\tau_2}$ &
 $\epsilon_{E_2}$ &
 $\sum_i \text{NT}_{i}\epsilon_{i}$&
 NT$\epsilon_{\text{Tot}}$ &
 Total candidates & S  & $\tau$ 
 \\
  &  &
  &
  &
  &
  &
 ($10^{24}$ atom yr)&($10^{24}$ atom yr)
  &Observed
  & (counts) & ($10^{24}$ yr) 
  \\\hline\hline
Semi-inclusive Visible\\\hline  
$^{76}\text{Ge}(ppp)\rightarrow ~^{73}\text{Cu}~e^{+}\pi^{+}\pi^{+}$ & 0.998 & 0.969 & 0.996 & N.A. & N.A. & 465.1 & 448.0 & 0 & 2.44 & $>$183.6 \\
$^{76}\text{Ge}(ppn)\rightarrow ~^{73}\text{Zn}~e^{+}\pi^{+}$ & 0.999 & 0.969 & 0.990 & N.A. & N.A. & 465.1 & 445.8 & 0 & 2.44 & $>$182.7\\
$^{76}\text{Ge}(pp)\rightarrow ~^{74}\text{Zn}~\pi^{+}\pi^{+}$ & 0.994 & 0.968 & 0.972 & N.A. & N.A. & 465.1 & 435.0 & 0 & 2.44 & $>$178.3 \\
$^{76}\text{Ge}(pn)\rightarrow ~^{74}\text{Ga}~\pi^{0}\pi^{+}$ & 0.979& 0.964 & 0.991 & N.A. & N.A. & 465.1 & 435.0 & 1 & 4.36 & $>$99.8\\
  &  &
  &
  &
  &
  &
 &
  &
  & & 
\\
%\hline
$^{74}\text{Ge}(ppp)\rightarrow ~^{71}\text{Cu}~e^{+}\pi^{+}\pi^{+}$ & 0.998 & 0.969 & 0.993 & N.A. & N.A. & 147.2 & 141.3 & 0 & 2.44 & $>$57.9\\
$^{74}\text{Ge}(ppn)\rightarrow ~^{71}\text{Zn}~e^{+}\pi^{+}$ & 0.999 & 0.967 & 0.982 & N.A. & N.A. & 147.2 & 139.6 & 0 & 2.44 & $>$57.2\\
  &  &
  &
  &
  &
  &
 &
  &
  & & 
\\
%\hline
$^{73}\text{Ge}(ppp)\rightarrow ~^{70}\text{Cu}~e^{+}\pi^{+}\pi^{+}$ & 0.998 & 0.968 & 0.996 & N.A. & N.A. & 17.6& 16.9& 0 & 2.44 & $>$6.9\\
$^{73}\text{Ge}(pnn)\rightarrow ~^{70}\text{Ga}~e^{+}\pi^{0}$ & 0.999 & 0.958 & 0.867 & N.A. & N.A. & 17.6 & 14.6 & 1 & 4.36& $>$3.3\\
$^{73}\text{Ge}(pp)\rightarrow ~^{71}\text{Zn}~\pi^{+}\pi^{+}$ & 0.994 & 0.967 & 0.982 & N.A. & N.A. & 17.6 & 16.6 & 0 & 2.44 & $>$6.8\\
$^{72}\text{Ge}(ppp)\rightarrow ~^{69}\text{Cu}~e^{+}\pi^{+}\pi^{+}$ & 0.998 & 0.967 & 0.973 & N.A. & N.A. & 62.1 & 58.4 & 0 & 2.44 & $>$23.9\\
$^{72}\text{Ge}(pn)\rightarrow ~^{70}\text{Ga}~\pi^{0}\pi^{+}$ & 0.979 & 0.958 & 0.867 & N.A. & N.A. & 62.1 & 50.5 & 1 & 4.36 & $>$11.6\\
$^{70}\text{Ge}(nnn)\rightarrow ~^{67}\text{Ge}~\bar{\nu}\pi^{0}$ & 0.952 & 0.959 & 0.972 & N.A. & N.A. & 46.5 & 41.3 & 1 & 4.36 & $>$9.5\\
\hline \hline   
Fully Inclusive\\  \hline
$^{76}\text{Ge}(ppp)\rightarrow ~^{73}\text{Cu}+\text{X}$ & N.A.& N.A. & 0.445 & 0.969 & 0.350 &343.5& 51.84 & 0 & 2.44 & $>$21.2 \\
$^{74}\text{Ge}(ppp)\rightarrow ~^{71}\text{Cu}+\text{X}$ & N.A. & N.A. & 0.114 & 0.969 & 0.050 & 109.0& 0.60 & 0 & 2.44 & $>$0.2 \\\hline
\end{tabular}
\caption{Efficiencies, exposures (enriched and natural detectors), signal upper limits, and partial lifetime limits for the modes of nucleon decay for the germanium isotopes for which the \DEM~has an interesting sensitivity. The signal upper limit (S) corresponds to Feldman-Cousins~\cite{Feldman:1997qc} 90\% confidence level interval for the Poisson signal for total candidates observed. ``N.A." is an abbreviation for ``not applicable". The index $i$ loops over PPC, ICPC, and BEGe, respectively. $\epsilon_{i}$ is $\epsilon_{\text{DC},i}$ for semi-inclusive visible modes and $\epsilon_{\text{PSD},i}^2$ for fully inclusive modes. $\text{NT}_{i}$ ($10^{24}$ atom yr) is the isotope exposure. $\text{NT}\epsilon_{\text{Tot}}$ is $(\sum_i \text{NT}_{i}\epsilon_{i})\times \epsilon_0\epsilon_{\tau_1}\epsilon_{E_1}$ for semi-inclusive visible modes and $(\sum_i \text{NT}_{i}\epsilon_{i})\times \epsilon_{E_1}\epsilon_{\tau_2}\epsilon_{E_2}$ for fully inclusive modes. }
\label{tab:lifetime}
\end{table*}

%\section{Event Selection}
The \DEM~consisted of germanium isotopes, with \76Ge~being the dominant isotope. The tri-nucleon decays of \76Ge include the following processes: \76Ge$(ppp)\rightarrow {}^{73}$Cu+X, \76Ge$(ppn)\rightarrow {}^{73}$Zn+X, \76Ge$(pnn) \rightarrow {}^{73}$Ga+X, and \76Ge$(nnn)\rightarrow {}^{73}$Ge+X where $\text{X}$ denotes either visible particles ($\text{X}_{vis}$) that deposit energy in the detectors or invisible particles which escape detection. If the visible particles, such as positrons or pions, deposit large amounts of energy in the detectors, we apply the semi-inclusive visible mode analysis~\footnote{The semi-inclusive visible mode is called the decay-specific mode in Ref.~\cite{Majorana:2018pdo}.}. Since the \DEM~cannot directly distinguish the $\text{X}_{vis}$ specific to each decay mode, it instead observes only the resulting energy deposits. Additionally, tri-nucleon decays may produce daughter isotopes in excited states, leading to the prompt emission of $\gamma$ rays. These $\gamma$ rays may either remain within the same detector or escape to be detected by another detectors, which can be identified through detector hit multiplicity in this scenario. On the other hand, if the emitted particles are invisible -- such as dark matter or neutrinos, where emissions escape from the detectors without depositing energy~\cite{Heeck:2019kgr} -- or if they are visible but have energies too low to saturate the detectors such as low-energy $\gamma$ rays, we apply the fully inclusive mode analysis~\footnote{The fully inclusive mode is called the invisible mode in Ref.~\cite{Majorana:2018pdo}.}. The summary of decay channels investigated for the isotopes $^{76}$Ge, $^{74}$Ge, $^{73}$Ge, $^{72}$Ge, and $^{70}$Ge can be found in Table~\ref{tab:lifetime}. 

The semi-inclusive visible and fully inclusive modes have distinct event detection sequences, but both share in the  production and detection of unstable daughter isotopes. In the fully inclusive mode, we observe a sequence of $\beta$ decays in unstable daughter isotopes. Conversely, the semi-inclusive visible modes involve detecting the release of substantial energy during the initial tri-nucleon decays, followed by the subsequent decay of an unstable daughter isotope. The decay spectra of $^{73}$Cu and $^{73}$Zn are presented in Ref.~\cite{Majorana:2018pdo}.

This study does not consider particle decay (proton or neutron emission) of the daughter nucleus resulting from baryon decay. For example, \76Ge can decay to the excited state of $^{74}$Zn via a $\Delta B=2$ process. The excited $^{74}$Zn can then emit a proton and decay to $^{73}$Cu. As a result, the detection of $^{73}$Cu does not necessarily imply a $\Delta B=3$ process. 

We adopt a procedure similar to the one employed in the previous study by the \DEM~\cite{Majorana:2018pdo}. For the semi-inclusive visible modes, the initial tri-nucleon decay must release a significant amount of energy, potentially saturating the detectors in the \DEM~system with a dynamic range set to approximately 10 MeV. It can trigger one or multiple detectors. It is worth noting that since the tri-nucleon decay occurs within a germanium detector, this energy release does not trigger the muon veto system. Consequently, for the semi-inclusive visible modes, the first triggered event consists of saturated events that are not tagged by the muon veto. The efficiency of these saturated events ($\epsilon_0$), as shown in Table~\ref{tab:lifetime}, was determined to be higher than 99$\%$ based on Monte Carlo simulation of high-energy $\pi$ or $e^+$ particles within the detector, which are generated according to the phase-space distributions of various decay channels, as detailed in Ref.~\cite{Majorana:2018pdo}. After the release of this substantial energy, the reaction leaves behind an unstable daughter isotope. For instance, ${}^{76}$Ge might decay into ${}^{73}$Cu through three-proton decays. The unstable daughter isotope ${}^{73}$Cu has a half-life of 4.2~s via $\beta$ decay and a $Q$ value of 6.6 MeV~\cite{ENSDF}. Additional relevant isotopes for this study are listed in Table~\ref{tab:isotope}. Since most saturated events are associated with muons or noise, a saturated event not related to these sources is exceedingly rare, resulting in a low background. Using such rare saturated events as a trigger to search for corresponding daughter isotope decays provides a distinct signature in the search for tri-nucleon decays, as it minimizes background interference.

The detection approach is as follows: we identify the first saturated, non-muon-veto event by applying the muon cut (removing muon-veto-tagged events). Within a specific time window, which is 5 times the half-life of the corresponding daughter isotope, we look for the presence of another event, which indicates a potential tri-nucleon decay. As detailed in Table~\ref{tab:isotope}, the longest half-life is 1268.4~s for ${}^{70}$Ga decay, resulting in a maximum time window of 6342~s. The corresponding efficiency for detecting the decay in a window of $5 T_{1/2}$ is given as $\epsilon_{\tau_1}$ in Table~\ref{tab:lifetime}. Some runs had significant gaps before the next run due to calibrations or other interruptions, such as detector failures or power outages. As a result, the efficiency was adjusted accordingly, similar to our previous study~\cite{Majorana:2018pdo}. To reduce the impact of low-energy noise and background, we search for events with energies between 100 keV and the $Q$ value of the corresponding isotope for the subsequent event from the daughter isotope. The associated efficiency of the energy cut ($\epsilon_{E_1}$ where $Q >E_1>100$ keV) is determined through Monte Carlo simulation~\cite{Boswell:2010mr, Majorana:2018pdo}. 

To reduce the noise and background events, we implement several cuts for the first trigger event, including the liquid nitrogen cut (which excludes events during the liquid nitrogen filling)~\cite{Majorana:2022udl}. Additionally, the good detector cut, which ensures that only well-functioning detectors are used, is applied universally to all events, regardless of noise or background considerations. For the subsequent event, we also apply the muon cut, the liquid nitrogen cut and the good detector cut. We permit germanium detector multiplicity greater than 1 in the second event, as the beta decay of the daughter isotope, which occurs within the same detector as the first event, can emit photons. Therefore, we require at least one hit in the same detector to ensure the detection of the beta decay from the daughter isotope, while additional hits in nearby detectors can result from emitted photons. These cuts have been considered within the context of active exposure, denoted as NT, where N represents the number of isotopic atoms within the active volume of the detectors, and T stands for the live time in years. The corresponding active exposure (NT$_i$) for different types of detectors, including PPC, ICPC and BEGe, are summarized in Table III. Furthermore, we apply multiple data cleaning (DC) cuts to discard nonphysical waveforms, pileup waveforms, and pulser events, extending the filtering process to the second event. The corresponding efficiencies ($\epsilon_{\text{DC},i}$ ) are summarized in Table III~\cite{Majorana:2022udl}.

For the fully inclusive modes, there may not be a saturated trigger event to indicate the reactions, so we employ a sequential search for $\beta$ decays of the daughter isotopes. For example, in the decay channel of ${}^{76}$Ge to ${}^{73}$Cu via $ppp$ tri-nucleon decay, the first trigger event arises from the $\beta$ decay of ${}^{73}$Cu (the daughter isotope) to ${}^{73}$Zn. After a designated time interval of 5 times the half-life of ${}^{73}$Zn, ${}^{73}$Zn (the granddaughter isotope) may undergo a $\beta$ decay to ${}^{73}$Ga. Due to the higher background associated with sequential $\beta$ decays as a trigger compared to a saturated event as a trigger, we seek events with energies between 2 MeV and the $Q$ value of the corresponding isotope for both the first daughter event and the second granddaughter event. The corresponding efficiencies of the energy cuts ($\epsilon_{E_1}$ and $\epsilon_{E_2}$ where $Q>E_1> 2$ MeV and $Q>E_2>2$ MeV) are also determined through Monte Carlo simulation. The time window is similarly set to 5 times the half-life of the corresponding granddaughter isotope, and the associated efficiency is denoted as $\epsilon_{\tau_2}$, as outlined in Table~\ref{tab:lifetime}. Together with the DC cuts, the muon cut, the liquid nitrogen cut and the good detector cut are also applied to both the first and second events. We permit germanium detector multiplicity greater than 1 but demand that at least one hit in each of the first and second events occur in the same detector. The decays of daughter (and granddaughter) isotopes are mainly $\beta$ decay in the bulk of the germanium detectors, occasionally accompanied by the emission of $\gamma$ rays, which may either be contained within or escape from the detector. To eliminate potential $\gamma$ and $\alpha$ background as well as surface events, pulse-shape discrimination (PSD) is applied alongside these DC cuts. This includes multiple-site rejection (AvsE), which primarily removes $\gamma$ scattering~\cite{Majorana:2019ftu}, delayed charge recovery (DCR), which mainly rejects surface $\alpha$ contamination~\cite{Majorana:2020xvk}, and late charge (LQ), where partial charge deposition in the transition layer between the n-type surfaces and the detector bulk leads to energy degradation~\cite{Majorana:2022udl}. The AvsE cut will also remove events involving $\beta$ decay with $\gamma$ emission contained within the same detector, but if the $\gamma$ escapes from the detector, the event will be retained. The efficiency of pure $\beta$ events without $\gamma$ emission in the same detector, subject to the energy cut, is summarized as $\epsilon_{E_i}$ in Table~\ref{tab:lifetime}. These events are selected individually in the Monte Carlo simulation, and the corresponding efficiency is then calculated. The corresponding efficiencies $\epsilon_{\text{PSD},i}$, including DC, AvsE, DCR, and LQ cuts, for the three detector types were estimated from experimental calibration data, as summarized in Table III. It is worth noting that we apply PSD to both the initial trigger event and the subsequent second event, resulting in an efficiency of $\epsilon_{\text{PSD},i}^2$. 

We found one potential sequence of events for the semi-inclusive visible modes search and one potential sequence of events for the fully inclusive modes passing the selection above. The semi-inclusive visible mode sequence occurred in a BEGe detector. The energy of the first event is $>$11558 keV and the energy of the second event is 147 keV. The time difference between the first event and the second event is 1712~s. It therefore satisfies as a candidate for the decay patterns of $^{76}\text{Ge}(pn)\rightarrow ~^{74}\text{Ga}~\pi^{0}\pi^{+}$, $^{73}\text{Ge}(pnn)\rightarrow ~^{70}\text{Ga}~e^{+}\pi^{0}$, $^{72}\text{Ge}(pn)\rightarrow ~^{70}\text{Ga}~\pi^{0}\pi^{+}$ and $^{70}\text{Ge}(nnn)\rightarrow ~^{67}\text{Ge}~\bar{\nu}\pi^{0}$ in the semi-inclusive visible modes. The fully inclusive mode sequence occurred in a BEGe detector. The energy of the first event is 2294 keV and the second event is 2614 keV, which is likely attributed to background from $^{208}$Tl decay. The time difference between the first event and the second event is 1657~s, which is too long to satisfy as a candidate for any decay pattern in the search channels of the fully inclusive modes. 
\begin{table}[ht]
    \centering
    \begin{tabular}{c|c|c}
        \hline
        Isotope & $T_{1/2}$ (s) & $Q$ (MeV) \\ \hline\hline
        $^{73}$Cu & 4.2 & 6.6 \\
        $^{73}$Zn & 24.5 & 4.1 \\
        $^{74}$Zn & 95.6 & 2.3 \\
        $^{74}$Ga & 487.2 & 5.4 \\
        $^{71}$Cu & 19.4 & 4.6 \\
        $^{71}$Zn & 147 & 2.8 \\
        $^{70}$Cu & 44.5 & 6.6 \\
        $^{70}$Ga & 1268.4 & 1.7 \\
        $^{69}$Cu & 171 & 2.7 \\
        $^{67}$Ge & 1134 & 4.2 \\
        \hline \hline     
    \end{tabular}
    \caption{The corresponding isotopes in Table~\ref{tab:lifetime} with their half-life and $Q$ value according to the ENSDF database~\cite{ENSDF}.}
    \label{tab:isotope}
\end{table}

%\section{Life Time Limits}

The partial lifetime limit ($\tau$) is calculated as follows
\begin{align}
\tau > \frac{\text{NT}\epsilon_{\text{Tot}}}{\text{S}}
\end{align}
where NT is the exposure and $\epsilon_{\text{Tot}}$ is the total efficiency. To convert the active exposure NT$_i$ from units of [kg yr] to [atom yr] as used in Table~\ref{tab:lifetime}, we divide it by the atomic mass $A_{i,j}$ (in g/mol) and then multiply by Avogadro's constant, $N_A=6.022\times 10^{23}$, while taking into account the respective fractional abundance of the isotope $f_{i,j}$. The conversion is given by:
\begin{align}
\text{NT}_i~[\text{atom yr}] = \sum_j\frac{\text{NT}_i [\text{kg yr}]\times 10^{3} \times N_A \times f_{i,j}}{A_{i,j}}
\end{align}
where the factor of $10^{3}$ converts the active mass from kg to g. Here, the index $i$ refers to different detector types (PPC, ICPC, and BEGe), and the index $j$ refers to the relevant isotopes.

The signal upper limit (S) corresponds to Feldman-Cousins~\cite{Feldman:1997qc} 90\% confidence level interval for the Poisson signal for total candidates observed. Table~\ref{tab:lifetime} shows the partial lifetime limit and corresponding parameters for different channels using the exposure of the complete data set with both enriched and natural detectors. 

For the semi-inclusive visible modes, we only consider the DC cuts to the subsequent second event. The $\text{NT}\epsilon_{\text{Tot}}$ is
\begin{align}
\text{NT}\epsilon_{\text{Tot}} = \left(\sum_{i} \text{NT}_{i}\epsilon_{\text{DC},i}\right)\epsilon_{0}\epsilon_{\tau_1}\epsilon_{E_1}
\end{align}
where the index $i$ loops over three types of detectors, PPC, ICPC, and BEGe. For the fully inclusive modes, we need to consider all PSD cuts to the initial trigger event and the subsequent second event. The $\text{NT}\epsilon_{\text{Tot}}$ becomes 
\begin{align}
\text{NT}\epsilon_{\text{Tot}} = \left(\sum_{i} \text{NT}_{i}\epsilon_{\text{PSD},i}^2\right)\epsilon_{E_1}\epsilon_{\tau_2}\epsilon_{E_2}
\end{align}
where the index $i$ also loops over PPC, ICPC, and BEGe.

With the complete data set from \DEM~and applying pulse-shape cuts (AvsE and LQ) to reduce background~\footnote{There were two sequence (Sequence 1 and 2) as shown in Table II of Ref.~\cite{Majorana:2018pdo}. In this study, the Sequence 1 was excluded due to AvsE and LQ cuts, while the Sequence 2 was excluded based on the AvsE cut. The new sequence introduced in this study is part of the updated dataset.}, we enhance the previous measurements~\cite{Majorana:2018pdo} of tri-nucleon decays in germanium isotopes, establishing new partial lifetime limits at $1.83\times10^{26}$ yr (90$\%$  confidence level) for \76Ge($ppp$) $\rightarrow{}^{73}$Cu e$^+\pi^+\pi^+$ and \76Ge($ppn$) $\rightarrow{}^{73}$Zn e$^+\pi^+$.

\begin{table}
\label{tab:Parameters}
\begin{center}
    \begin{tabular}{c|c|c|c}
	\hline 
	\hline
$i$& PPC & ICPC & BEGe\\ \hline\hline
$\text{NT}_{i}$ & $61.64^{+0.89}_{-1.17}$ kg yr & $2.82^{+0.04}_{-0.05}$ kg yr & $27.383$ kg yr\\
%$^{76}$Ge Enrichment & $87.4\pm 0.5\%$ & $88.0\pm1.0\%$ & $7.75\%$\\
$\epsilon_{\text{DC},i}$ & $99.9\pm 0.1\%$ & $99.9\pm 0.1\%$ & $99.9\pm 0.1\%$\\
$\epsilon_{\text{PSD},i}$ & $86.1\pm 3.9\%$ & $81.0^{+5.3}_{-7.3}\%$ & $86.1\pm 3.9\%$\\
\hline \hline     
\end{tabular}
\end{center}
\caption{A summary of key analysis parameters~\cite{Majorana:2022udl}. For the semi-inclusive visible modes, the only PSD cut is the data cleaning ($\epsilon_{\text{DC},i}$). For the fully inclusive mode, all PSD cuts are used including DC, AvsE, DCR, and LQ cuts ($\epsilon_{\text{PSD},i}$).}
\end{table}

%\section{Discussion}

The systematic uncertainties encompass several components, including the exposure uncertainty (\mbox{$<2\%$}), uncertainty associated with the DC cuts (\mbox{$<0.1\%$}), uncertainty related to the PSD cuts (\mbox{$<5\%$}), uncertainty arising from the simulation model (\mbox{$2\%$}), and the statistical uncertainty of the simulated efficiencies (\mbox{$<1\%$}). Since the limit is primarily constrained by statistical considerations, we ignored their contribution and continued to employ the Feldman-Cousins technique~\cite{Feldman:1997qc}.

For the fully inclusive model, we applied all pulse-shape cuts used in the \0vbb analysis~\cite{Majorana:2022udl}, including DCR, AvsE, and LQ. In contrast, the previous study~\cite{Majorana:2018pdo} in the \DEM~used only the DCR pulse-shape cut to improve efficiency, as there were fewer background events in that dataset. To enhance the signal-to-background ratio in this study, we introduced the additional pulse-shape cuts (AvsE and LQ). The corresponding efficiency was determined via Monte Carlo simulations, where events with coincident $\gamma$ emission alongside $\beta$ decay were excluded.

Additional background in the semi-inclusive visible modes arises from muon-induced saturated events that were not properly tagged by the veto system and from random time coincidences. For random time coincidences, the event rate for events with energy greater than 100 keV in the \DEM~is approximately $10^{-4}$ Hz. After a saturated event, the expected number of random events is calculated by multiplying the longest time window by the event rate, yielding $6342 \times 10^{-4} \approx 0.6$. Additionally, the subsequent event must occur in the same detector as the trigger-saturated event. Given that the \DEM~has around 35 detectors (with variations across different datasets), the background from saturated, non-muon-veto events and random time coincidences is approximately 0.017 (0.6/35) counts. This background remains low enough to preserve our discovery potential or to set a new lower limit on the partial lifetime for rare multi-nucleon decays. Although the muon detection efficiency is excellent~\cite{Majorana:2016ifg} with near-$4\pi$ geometric coverage, six out of 492 saturated events were not tagged by the veto system. Of these, two are likely non-physical events, possibly caused by an electrical breakdown (as discussed in Ref.~\cite{Majorana:2018pdo}). Three of these events lack corresponding daughter events within the time window due to the low event rate. However, one saturated, non-muon-veto event satisfies the criteria defined in this study. If more such events are found and their time difference distribution aligns with the half-life of the corresponding daughter isotope's $\beta$ decay, it would strongly enhance the likelihood of confirming the rare nucleon decays.

The GERDA experiment recently established partial lifetime limits of approximately $1.2\times 10^{26}$ years for fully inclusive tri-nucleon decays of \76Ge with an exposure of 61.89 kg yr~\cite{GERDA:2023uuw}.  GERDA developed an algorithm to identify the two-step waveform from the metastable state $^{73m}$Ge decay but did not observe any such events. In contrast, the \DEM~has observed several events consistent with $^{73m}$Ge decay, which is believed to result from the decay of $^{73}$As due to cosmogenic background radiation when the detectors were on the surface~\cite{Majorana:2021lgr}. Due to the high number of $^{73m}$Ge events observed in the \DEM, our sensitivity to identify tri-nucleon decays using the same algorithm as GERDA is significantly reduced.

 %\section{Conclusion}
In this study, the \DEM~achieves the most stringent partial lifetime limit for the process of \76Ge($ppp$) $\rightarrow$ $^{73}$Cu e$^+\pi^+\pi^+$ and \76Ge($ppn$) $\rightarrow$ $^{73}$Zn e$^+\pi^+$ to date at $1.83\times 10^{26}$ years. LEGEND-1000 (Large Enriched Germanium Experiment for Neutrinoless $\beta\beta$ Decay)~\cite{LEGEND:2021bnm}, the forthcoming generation of ton-scale germanium experiment, aims to reach an exposure of approximately $10^4$ kg yr. It is anticipated that LEGEND will significantly improve the current tri-nucleon decay limits by at least two orders of magnitude, potentially extending the limit to $10^{28}$ yr and beyond.

%\section*{Acknowledgments}

This material is based upon work supported by the U.S.~Department of Energy, Office of Science, Office of Nuclear Physics under contract / award numbers DE-AC02-05CH11231, DE-AC05-00OR22725, DE-AC05-76RL0130, DE-FG02-97ER41020, DE-FG02-97ER41033, DE-FG02-97ER41041, DE-SC0012612, DE-SC0014445, DE-SC0017594, DE-SC0018060, DE-SC0022339, and LANLEM77/LANLEM78. We acknowledge support from the Particle Astrophysics Program and Nuclear Physics Program of the National Science Foundation through grant numbers MRI-0923142, PHY-1003399, PHY-1102292, PHY-1206314, PHY-1614611, PHY-13407204, PHY-1812409, 
and PHY-2209530. 
We gratefully acknowledge the support of the Laboratory Directed Research \& Development (LDRD) program at Lawrence Berkeley National Laboratory for this work. We gratefully acknowledge the support of the U.S.~Department of Energy through the Los Alamos National Laboratory LDRD Program, the Oak Ridge National Laboratory LDRD Program, and the Pacific Northwest National Laboratory LDRD Program for this work.  
This research used resources provided by the Oak Ridge Leadership Computing Facility at Oak Ridge National Laboratory and by the National Energy Research Scientific Computing Center, a U.S.~Department of Energy Office of Science User Facility. We thank our hosts and colleagues at the Sanford Underground Research Facility for their support. 

{\it Data availability}. The data that support the findings of this article are openly available~\cite{Majorana:2023kmv}, embargo periods may apply.

\bibliographystyle{unsrt}

%\bibliography{main}% Produces the bibliography via BibTeX.

\begin{thebibliography}{10}

\bibitem{Sakharov:1967dj}
A.~D. Sakharov.
\newblock {Violation of CP Invariance, C asymmetry, and baryon asymmetry of the
  universe},
\newblock {\bf Pisma Zh. Eksp. Teor. Fiz.}, 5:32--35, 1967.
\newblock{[{\href{https://doi.org/10.1070/PU1991v034n05ABEH002497}{Sov. Phys. Usp. 34, 392 (1991)}}]}

\bibitem{Babu:2013jba}
K.~S. Babu {\it et~al.},
\newblock {Baryon Number Violation},
\newblock {\href{https://doi.org/10.48550/arXiv.1311.5285}{arXiv:1311.5285}}.

\bibitem{Dev:2022jbf}
P.~S.~B. Dev {\it et~al.},
\newblock {Searches for Baryon Number Violation in Neutrino Experiments: A
  White Paper},
\newblock {\href{https://doi.org/10.1088/1361-6471/ad1658}{{\bf J. Phys. G: Nucl. Part. Phys.}, 51, 033001 (2024)}}.

\bibitem{FileviezPerez:2022ypk}
Pavel Fileviez~Perez {\it et~al.},
\newblock {On Baryon and Lepton Number Violation},
\newblock {\href{https://doi.org/10.48550/arXiv.2208.00010}{arXiv:2208.00010}}.

\bibitem{Super-Kamiokande:2016exg}
K.~Abe {\it et~al.},
\newblock {Search for proton decay via $p \to e^+\pi^0$ and $p \to \mu^+\pi^0$
  in 0.31 megaton\textperiodcentered{}years exposure of the Super-Kamiokande
  water Cherenkov detector},
\newblock {\href{https://doi.org/10.1103/PhysRevD.95.012004}{{\bf Phys. Rev. D}, 95(1):012004, 2017}}.

\bibitem{Super-Kamiokande:2015pys}
V.~Takhistov {\it et~al.},
\newblock {Search for Nucleon and Dinucleon Decays with an Invisible Particle
  and a Charged Lepton in the Final State at the Super-Kamiokande Experiment},
\newblock {\href{https://doi.org/10.1103/PhysRevLett.115.121803}{{\bf Phys. Rev. Lett.}, 115(12):121803, 2015}}.

\bibitem{Super-Kamiokande:2015jbb}
J.~Gustafson {\it et~al.},
\newblock {Search for dinucleon decay into pions at Super-Kamiokande},
\newblock {\href{https://doi.org/10.1103/PhysRevD.91.072009}{{\bf Phys. Rev. D}, 91(7):072009, 2015}}.

\bibitem{Babu:2003qh}
K.~S. Babu, Ilia Gogoladze, and Kai Wang.
\newblock {Gauged baryon parity and nucleon stability},
\newblock {\href{https://doi.org/10.1016/j.physletb.2003.07.036}{{\bf Phys. Lett. B}, 570:32--38, 2003}}.

\bibitem{Bakker:2004}
B.L.G. Bakker, A.I. Veselov, and M.A. Zubkov.
\newblock A hidden symmetry in the standard model,
\newblock {\href{https://doi.org/10.1016/j.physletb.2003.12.062}{{\bf Physics Letters B}, 583(3–4):379–382, March 2004}}.

\bibitem{Bernabei:2006tw}
R.~Bernabei {\it et~al.},
\newblock {Search for rare processes with DAMA/LXe experiment at Gran Sasso},
\newblock {\href{https://doi.org/10.1140/epja/i2006-08-004-y}{{\bf Eur. Phys. J. A}, 27(S1):35--41, 2006}}.

\bibitem{EXO-200:2017hwz}
J.~B. Albert {\it et~al.},
\newblock {Search for nucleon decays with EXO-200},
\newblock {\href{https://doi.org/10.1103/PhysRevD.97.072007}{{\bf Phys. Rev. D}, 97(7):072007, 2018}}.

\bibitem{Hazama:1994zz}
R.~Hazama, H.~Ejiri, K.~Fushimi, and H.~Ohsumi.
\newblock {Limits on single- and multinucleon decays in I-127 by inclusive
  measurements of nuclear gamma and x rays},
\newblock {\href{https://doi.org/10.1103/PhysRevC.49.2407}{{\bf Phys. Rev. C}, 49:2407--2412, 1994}}.

\bibitem{Heeck:2019kgr}
Julian Heeck and Volodymyr Takhistov.
\newblock {Inclusive Nucleon Decay Searches as a Frontier of Baryon Number Violation},
\newblock {\href{https://doi.org/10.1103/PhysRevD.101.015005}{{\bf Phys. Rev. D}, 101(1):015005, 2020}}.

\bibitem{Majorana:2018pdo}
S.~I. Alvis {\it et~al.},
\newblock {Search for trinucleon decay in the {\sc Majorana Demonstrator}},
\newblock {\href{https://doi.org/10.1103/PhysRevD.99.072004}{{\bf Phys. Rev. D}, 99(7):072004, 2019}}.

\bibitem{GERDA:2023uuw}
M.~Agostini {\it et~al.},
\newblock {Search for tri-nucleon decays of $^{76}$Ge in GERDA},
\newblock {\href{https://doi.org/10.1140/epjc/s10052-023-11862-8}{{\bf Eur. Phys. J. C}, 83(9):778, 2023}}.

\bibitem{Majorana:2021lgr}
I.~J. Arnquist {\it et~al.},
\newblock {Signatures of muonic activation in the {\sc Majorana Demonstrator}},
\newblock {\href{https://doi.org/10.1103/PhysRevC.105.014617}{{\bf Phys. Rev. C}, 105(1):014617, 2022}}.

\bibitem{Proceedings:2020nzz}
K.~S. Babu {\it et~al.},
\newblock { $|\Delta \mathcal{B}| =2$: A State of the Field, and Looking
  Forward--A brief status report of theoretical and experimental physics
  opportunities},
\newblock {\href{https://doi.org/10.48550/arXiv.2010.02299}{{arXiv:2010.02299}}}.

\bibitem{Majorana:2013cem}
N.~Abgrall {\it et~al.},
\newblock {The {\sc Majorana Demonstrator}~Neutrinoless Double-Beta Decay
  Experiment},
\newblock {\href{ https://doi.org/10.1155/2014/365432}{{\bf Adv. High Energy Phys.}, 2014:365432, 2014}}.

\bibitem{Abgrall:2025tsj}
N.~Abgrall {\it et~al.},
\newblock{The {\sc Majorana Demonstrator} experiment's construction, commissioning, and performance},
 \newblock{\href{https://doi.org/10.48550/arXiv.2501.02060}{{arXiv:2501.02060}}}.

\bibitem{Heise:2022iaf}
Jaret Heise.
\newblock {The Sanford Underground Research Facility},
\newblock {\href{https://doi.org/10.48550/arXiv.2203.08293}{arXiv.2203.08293}}.

\bibitem{Barbeau:2007qi}
P.~S. Barbeau, J.~I. Collar, and O.~Tench.
\newblock {Large-Mass Ultra-Low Noise Germanium Detectors: Performance and
  Applications in Neutrino and Astroparticle Physics},
\newblock {\href{https://doi.org/10.1088/1475-7516/2007/09/009}{{\bf J. Cosmol. Astropart. Phys.}, 09:009, 2007}}.

\bibitem{Majorana:2016ifg}
N.~Abgrall {\it et~al.},
\newblock {Muon Flux Measurements at the Davis Campus of the Sanford
  Underground Research Facility with the {\sc Majorana Demonstrator}~Veto
  System},
\newblock {\href{https://doi.org/10.1016/j.astropartphys.2017.01.013}{{\bf Astropart. Phys.}, 93:70--75, 2017}}.

\bibitem{Hoppe:2014nva}
E.~W. Hoppe, C.~E. Aalseth, O.~T. Farmer, T.~W. Hossbach, M.~Liezers, H.~S.
  Miley, N.~R. Overman, and J.~H. Reeves.
\newblock {Reduction of radioactive backgrounds in electroformed copper for
  ultra-sensitive radiation detectors},
\newblock {\href{https://doi.org/10.1016/j.nima.2014.06.082}{{\bf Nucl. Instrum. Methods Phys. Res.  A}, 764:116--121, 2014}}.

\bibitem{Majorana:2016lsk}
N.~Abgrall {\it et~al.},
\newblock {The {\sc Majorana Demonstrator}~radioassay program},
\newblock {\href{https://doi.org/10.1016/j.nima.2016.04.070}{{\bf Nucl. Instrum. Methods Phys. Res. A}, 828:22--36, 2016}}.

\bibitem{Majorana:2021mtz}
N.~Abgrall {\it et~al.},
\newblock {The {\sc Majorana Demonstrator}~readout electronics system},
\newblock {\href{https://doi.org/10.1088/1748-0221/17/05/T05003}{{\bf J. Inst.}, 17(05):T05003, 2022}}.

\bibitem{Majorana:2019ftu}
S.~I. Alvis {\it et~al.},
\newblock {Multisite event discrimination for the {\sc Majorana Demonstrator}},
\newblock {\href{https://doi.org/10.1103/PhysRevC.99.065501}{{\bf Phys. Rev. C}, 99(6):065501, 2019}}.

\bibitem{Majorana:2020xvk}
I.~J. Arnquist {\it et~al.},
\newblock {$\alpha $-event characterization and rejection in point-contact HPGe
  detectors},
\newblock {\href{https://doi.org/10.1140/epjc/s10052-022-10161-y}{{\bf Eur. Phys. J. C}, 82(3):226, 2022}}.

\bibitem{Majorana:2020llj}
N.~Abgrall {\it et~al.},
\newblock {ADC Nonlinearity Correction for the {\sc Majorana Demonstrator}},
\newblock {\href{https://doi.org/10.1109/TNS.2020.3043671}{{\bf IEEE Trans. Nucl. Sci.}, 68(3):359--367, 2021}}.

\bibitem{Majorana:2022vai}
I.~J. Arnquist {\it et~al.},
\newblock {Charge trapping correction and energy performance of the {\sc
  Majorana Demonstrator}},
\newblock {\href{https://doi.org/10.1103/PhysRevC.107.045503}{{\bf Phys. Rev. C}, 107(4):045503, 2023}}.

\bibitem{Majorana:2017vdg}
N.~Abgrall {\it et~al.},
\newblock {The {\sc Majorana Demonstrator}~calibration system},
\newblock {\href{https://doi.org/10.1016/j.nima.2017.08.005}{{\bf Nucl. Instrum. Methods Phys. Res. A}, 872:16--22, 2017}}.

\bibitem{Majorana:2023kdx}
I.~J. Arnquist {\it et~al.},
\newblock {Energy calibration of germanium detectors for the {\sc Majorana
  Demonstrator}},
\newblock {\href{https://doi.org/10.1088/1748-0221/18/09/P09023}{{\bf J Inst.}, 18(09):P09023, 2023}}.

\bibitem{Luke:1989}
P.~N.~Luke, F.~S. Goulding, N.~W. Madden, and R.H. Pehl,
\newblock Low capacitance large volume shaped-field germanium detector,
\newblock {\href{https://doi.org/10.1109/23.34577}{{\bf IEEE Transactions on Nuclear Science}, 36(1):926--930, 1989}}.

\bibitem{COOPER201125}
R.~J.~Cooper, D.~C. Radford, P.~A. Hausladen, and K.~Lagergren,
\newblock A novel HPGe detector for gamma-ray tracking and imaging,
\newblock {\href{https://doi.org/10.1016/j.nima.2011.10.008}{{\bf Nucl. Instrum. Methods Phys. Res. A}, 665:25--32, 2011}}.

\bibitem{Canberra}
{Canberra Industries Inc. (now Mirion Technologies), 800 Research Parkway
  Meriden, CT 06450},
\newblock information available
  at~\url{https://www.mirion.com/products/bege-broad-energy-germanium-detectors},
  2023,
\newblock accessed on October 27, 2023.

\bibitem{Majorana:2022udl}
I.~J. Arnquist {\it et~al.},
\newblock {Final Result of the {\sc Majorana Demonstrator}\textquoteright{}s
  Search for Neutrinoless Double-\ensuremath{\beta} Decay in $^{76}$Ge},
\newblock {\href{https://doi.org/10.1103/PhysRevLett.130.062501}{{\bf Phys. Rev. Lett.}, 130(6):062501, 2023}}.

\bibitem{Note1}
The semi-inclusive visible mode is called the decay-specific mode in Ref.~\cite{Majorana:2018pdo}.

\bibitem{Note2}
The fully inclusive mode is called the invisible mode in Ref.~\cite{Majorana:2018pdo}. 

\bibitem{Feldman:1997qc}
Gary~J. Feldman and Robert~D. Cousins,
\newblock {Unified approach to the classical statistical analysis of small
  signals},
\newblock {\href{https://doi.org/10.1103/PhysRevD.57.3873}{{\bf Phys. Rev. D}, 57:3873--3889, 1998}}.

\bibitem{ENSDF}
{ENSDF Database},
\newblock version available at \url{http://www.nndc.bnl.gov}, 2023,
\newblock accessed on October 27, 2023.

\bibitem{Boswell:2010mr}
Melissa Boswell {\it et~al.},
\newblock {MaGe-a Geant4-based Monte Carlo Application Framework for
  Low-background Germanium Experiments}.
\newblock {\href{https://doi.org/10.1109/TNS.2011.2144619}{{\bf IEEE Trans. Nucl. Sci.}, 58:1212--1220, 2011}}.

\bibitem{Note3}
There were two sequences (Sequence 1 and 2) as shown in Table II of Ref.~\cite
  {Majorana:2018pdo}. In this study, Sequence 1 was excluded due to AvsE and LQ
  cuts, while Sequence 2 was excluded based on the AvsE cut. The new case
  introduced in this study is part of the updated dataset.

\bibitem{LEGEND:2021bnm}
N.~Abgrall {\it et~al.},
\newblock {The Large Enriched Germanium Experiment for Neutrinoless
  $\beta\beta$ Decay}: {LEGEND-1000 Preconceptual Design Report}, 7 2021. \href{https://doi.org/10.48550/arXiv.2107.11462}{arxiv:2107.11462}

\bibitem{Majorana:2023kmv}
I.~J. Arnquist {\it et~al.},
\newblock { {\sc Majorana Demonstrator} Data Release for AI/ML Applications},
\newblock {\href{https://doi.org/10.48550/arXiv.2308.10856}{arxiv:2308.10856}},
\newblock {\href{https://doi.org/10.5281/zenodo.8257027}{zenodo.8257027}}


\end{thebibliography}

\end{document}